\documentclass{elsart1p}
\usepackage{graphics}
\usepackage{graphicx}
\usepackage{bm}% bold math
\usepackage{epsfig}
\usepackage{amsmath,amssymb}

\usepackage[normalem]{ulem}  % \sout{old text} for strikeout
\usepackage{color} % For blue in-text comments and additions

\renewcommand\sout{\bgroup \color{red} \ULdepth=-.5ex \ULset}
\newcommand{\Slash}[1]{\ooalign{\hfil/\hfil\crcr$#1$}}

\begin{document}

\begin{frontmatter}

\title{Production of doubly charmed tetraquarks\\
with exotic color configurations\\
in electron-positron collisions}

%\author{Yoichi Ikeda}
% \corauthref{cor}},
%\corauth[cor]{Corresponding author.}
%\ead{yikeda@riken.jp}

\author[a]{Tetsuo Hyodo\corauthref{cor}},
\ead{hyodo@th.phys.titech.ac.jp}
\author[b]{Yan-Rui Liu},
\author[a,c]{Makoto Oka},
\author[d]{Kazutaka Sudoh}, 
\author[e]{Shigehiro Yasui}
\ead{yasuis@post.kek.jp}

\address[a]{Department of Physics, Tokyo Institute of Technology,
Tokyo 152-8551, Japan}
\address[b]{School of Physics, Shandong University, Jinan, 250100, PR China}

\address[c]{J-PARC Branch, KEK Theory Center, Institute of Particle and Nuclear Studies, High Energy Accelerator Research Organization (KEK), 203-1, Shirakata, Tokai, Ibaraki, 319-1106, Japan}

\address[d]{Nishogakusha University, 6-16, Sanbancho, Chiyoda,
Tokyo, 102-8336, Japan}

\address[e]{KEK Theory Center, Institute of Particle and Nuclear
Studies, High Energy Accelerator Research Organization, 1-1, Oho,
Ibaraki, 305-0801, Japan}

\corauth[cor]{Corresponding author at: Department of Physics, Tokyo Institute of Technology, Ookayama 2-12-1-H27, Meguro, Tokyo 152-8551, Japan}



\begin{abstract}
Structure and production of doubly charmed tetraquarks T$_{\rm cc}$ (cc$\bar{\rm u}\bar{\rm d}$) are studied from the viewpoint of color configurations. Based on the diquark correlation, the tetraquark T$_{\rm cc}$ with $I(J^{P})=0(1^{+})$ is considered to be stable against strong decay. We discuss that the mixing probability of color antitriplet and sextet cc components in T$_{\rm cc}$ is suppressed by $1/m_{\rm c}^{2}$, so the two configurations are separately realized in the heavy quark limit. Utilizing the nonrelativistic QCD framework, we evaluate the production cross sections of T$_{\rm cc}$ in electron-positron collisions. The momentum dependence of the cross section of color antitriplet is found to be different from that of sextet, which can be used to discriminate the color structure of the T$_{\rm cc}$ states in experimental measurements.\end{abstract}
\end{frontmatter}

Study of exotic hadrons is one of the most interesting subjects in the quark and hadron physics in these years. The experimental discoveries of the exotic hadrons, such as X, Y, Z$^{\pm}$ in the charm sector and Y$_{\rm b}$, Z$_{\rm b}^{\pm}$ in the bottom sector, have motivated many researchers toward the study of those heavy-flavor exotic hadrons \cite{Brambilla:2004wf,Swanson:2006st,Voloshin:2007dx,Nielsen:2009uh,Brambilla:2010cs}. The exotic hadrons have many unexpected properties, such as masses, decay widths, branching ratios, which are hard to be explained in the conventional quark model. The doubly charmed ($C=2$) tetraquark T$_{\rm cc}$ with four quark configuration cc$\bar{\rm u}\bar{\rm d}$ (two charm quarks c's and light up and down antiquarks $\bar{\rm u}$ and $\bar{\rm d}$) is a new interesting candidate, because it is a genuine tetraquark hadron.\footnote{In literature, another definition of T$_{\rm cc}$ as $\bar{\rm c}\bar{\rm c}$ud has also been used. Although we deal with the cc$\bar{\rm u}\bar{\rm d}$ state in this Letter, all the discussions can be equally applied to $\bar{\rm c}\bar{\rm c}$ud state.} The mass of T$_{\rm cc}$ as a tetraquark state has been theoretically studied from the quark model \cite{Ader:1981db,Zouzou:1986qh,Lipkin:1986dw,Heller:1986bt,Carlson:1987hh,SilvestreBrac:1993ss,SilvestreBrac:1993ry,Semay:1994ht,Pepin:1996id,SchaffnerBielich:1998ci,Brink:1998as,Janc:2004qn,Barnea:2006sd,Cohen:2006jg,Ebert:2007rn,Lee:2007tn,Vijande:2007ix,Vijande:2007fc,Vijande:2007rf,Zhang:2007mu,Lee:2009rt,Vijande:2009kj,Vijande:2009zs,Yang:2009zzp,Carames:2011zz,Vijande:2011zz} and from the QCD sum rules~\cite{Navarra:2007yw}. It is studied also as molecule-like DD, DD$^{\ast}$ and D$^{\ast}$D$^{\ast}$ bound or resonant states in hadronic molecule picture \cite{Manohar:1992nd,Tornqvist:1993ng,Ding:2009vj,Molina:2010tx,Ohkoda:2011vj}. One of the interesting properties of T$_{\rm cc}$ is its color configurations. A quark-quark pair in T$_{\rm cc}$ can be combined, not only in color antitriplet ($\bm{\bar{3}}$) configuration, but also in color sextet ($\bm{6}$) configuration, the latter of which does not exist in normal hadrons with qqq or q$\bar{\rm q}$. If T$_{\rm cc}$ exists, it helps us to understand the quark-quark interaction in the color channel which is not accessible in normal hadrons. To research the existence of T$_{\rm cc}$, it is required to search for T$_{\rm cc}$ by experiments in high energy accelerator facilities. The study in hadron-hadron collisions and heavy ion collisions, such as in Tevatron, RHIC and LHC, is discussed in Refs.~\cite{DelFabbro:2004ta,Cho:2010db,Cho:2011ew}. Moreover, analyses of double charm productions in electron-positron collisions are also carried out at Belle and BaBar~\cite{Abe:2002rb,Abe:2004ww,Aubert:2005tj}. In the present study, we consider T$_{\rm cc}$ as a tetraquark state and study how T$_{\rm cc}$ can be produced in electron-positron collisions.

In literature, there have been many fully dynamical quark model calculations of the mass of T$_{\rm cc}$. In shortly summarized, their results can be essentially understood by the simple diquark picture~\cite{Lee:2007tn,Lee:2009rt} as shown below. We consider that diquarks are strongly correlated pair of quarks through color-spin interaction~\cite{Jaffe:2004ph};
\begin{eqnarray}
H_{\rm int} = \sum_{i<j} \frac{C_{\rm H}}{m_i m_j} 
\left(-\frac{3}{8}\right)\vec{\lambda}_{i} \!\cdot\! \vec{\lambda}_{j} \, \vec{s}_{i} \!\cdot\! \vec{s}_{j}
.
\label{eq:H_int}
\end{eqnarray}Here $C_{\rm H}$ is the coupling strength, $m_{i}$ is the mass of the quark $i$, $\vec{\lambda}_{i}$ is the Gell-Mann matrix operating to the color of the quark $i$, and $\vec{s}_{i}=\vec{\sigma}_{i}/2$ is the spin operator with the Pauli matrix $\vec{\sigma}_{i}$ operating to the spin of the quark $i$. The factor $-3/8$ is multiplied for normalization ($-3/8 \vec{\lambda}_{i} \!\cdot\! \vec{\lambda}_{j} = 1$) for the pair of quarks $i$ and $j$ in color antitriplet ($\bm{\bar{3}}$) channel. The values of $\vec{\lambda}_{i} \!\cdot\! \vec{\lambda}_{j} \, \vec{s}_{i} \!\cdot\! \vec{s}_{j}$ are dependent on the color-spin channels as summarized in Table~\ref{table:color-spin}. We note $C_{\rm H}$ is given as $C_{\rm H}=v_{0} \langle \delta(r_{ij}) \rangle$, where $v_{0}$ is related to the interaction constant, $r_{ij}$ is the distance between the quarks $i$ and $j$, and the expectation value of the delta function is given from the wave function of the relative coordinate for the quarks $i$ and $j$. The parameters are given by $C_{\rm H}=C_{\rm B}$ ($2C_{\rm H}=C_{\rm M}$) for quark-quark (quark-antiquark) pair with $C_{\rm B}/m_{\rm u}^{2}=193$ MeV ($C_{\rm M}/m_{\rm u}^{2}=635$ MeV), $m_{\rm u}=m_{\rm d}=300$ MeV and $m_{\rm c}=1500$ MeV to reproduce the masses of known hadrons \cite{Lee:2007tn,Lee:2009rt}. Note that $C_{\rm M}$ is larger than $C_{\rm B}$, reflecting the color factor by two and the difference of the quark wave function. For example, in $\Lambda_{\rm c}$, there is an attractive ud diquark (``good" diquark) with color $\bm{\bar{3}}$, spin $^{1}{\rm S}_{0}$ in isospin $I=0$.\footnote{It is important to note that the attractive ud diquark is a source to form the color superconductivity in quark matter at high density~\cite{Alford:1998zt,Rapp:1997zu,Alford:2007xm}.} In $\Sigma_{\rm c}$ and $\Sigma_{\rm c}^{\ast}$, there is a repulsive ud diquark (``bad" diquark) with color $\bm{\bar{3}}$, but spin $^{3}{\rm S}_{1}$ in isospin $I=1$~\cite{Jaffe:2004ph}.

As a remark, we note that $C_{\rm B}$ for heavy-heavy (cc) quark pair should be different from those of light-light and light-heavy quark pairs. This is because the wave function of the heavy-heavy quark pair is spatially shrunk and its value at the center of mass is increased. Hence the expectation value $\langle \delta(r) \rangle$ by the wave function ($r$ the distance between the two quarks) is increased. Therefore, $C_{\rm H}$ for charm quark pairs should be different from the other cases as $C_{\rm H}=C_{\rm cc}$ with $C_{\rm cc}/m_{\rm c}^2=39$ MeV, which causes only the minor difference~\cite{Lee:2007tn,Lee:2009rt}.

\begin{table}[bp]
\caption{The expectation values of $(-3/8)\vec{\lambda}_{i} \!\cdot\! \vec{\lambda}_{j} \, \vec{s}_{i} \!\cdot\! \vec{s}_{j}$ for quarks $i$ and $j$ with spin $s=0$, 1 and color $\bm{\bar{3}}$, $\bm{6}$.}
\begin{center}
\begin{tabular}{crr}
 \hline
 \vspace{0.1cm}
  & $\bm{\bar{3}}$ & $\bm{6}$ \\
 \hline
 \vspace{0.1cm}
 $s=0$ & $-\dfrac{3}{4}$ & $\dfrac{3}{8}$ \\
 \vspace{0.1cm}$s=1$ & $\dfrac{1}{4}$ & $-\dfrac{1}{8}$ 
  \\
 \hline
\end{tabular}
\end{center}
\label{table:color-spin}
\end{table}%

% Tcc(3c)
Now let us see how T$_{\rm cc}$ can be a stable particle in the diquark model~\cite{Lee:2007tn,Lee:2009rt}. The flavor dependence of the color-spin interaction is important to the stability of T$_{\rm cc}$. We note that the color-spin interaction $H_{\rm int}$ is proportional to the quark mass factor $1/m_{i}m_{j}$. It indicates that the light quarks are affected much by the color-spin interaction, while the heavy quarks are not. We thus consider a state with ``good'' $\bar{\rm u}\bar{\rm d}$ diquark and cc diquark with color $\bm{\bar{3}}$ and spin $^{3}{\rm S}_{1}$. The repulsion of cc diquark is very small due to the suppression with the $1/m_{\rm c}^{2}$ factor, while the $\bar{\rm u}\bar{\rm d}$ diquark induces a strong attraction to make a bound state. In this picture, the quantum number of the ground state of T$_{\rm cc}$ is uniquely determined as $I(J^{P})=0(1^{+})$. The lowest two-meson threshold in this channel is the D and D$^{\ast}$ mesons in s wave. We can estimate the binding energy of T$_{\rm cc}$ from the DD$^{\ast}$ threshold as
\begin{align}
{\rm B.E.}&=  \left( -\frac{3}{4} \frac{C_{\rm M}}{m_{\rm u}m_{\rm c}} + \frac{1}{4} \frac{C_{\rm M}}{m_{\rm u}m_{\rm c}} \right)
-\left(- \frac{3}{4} \frac{C_{\rm B}}{m_{\rm u}^2} + \frac{1}{4} \frac{C_{\rm cc}}{m_{\rm c}^2} \right)
\simeq 71\hspace{0.5em} {\rm MeV},
\nonumber
\end{align}
where the first term is the sum of the energies from color-spin interaction in D and D$^{\ast}$ mesons and the second term is the energy from color-spin interaction in T$_{\rm cc}$. The above result means that the mass of T$_{\rm cc}$ is 71 MeV below the DD$^{\ast}$ threshold. Thus, T$_{\rm cc}$ can be a stable bound state as a genuine flavor exotic hadron.

% possibility of Tcc(6c)
We consider a new state of T$_{\rm cc}$ whose color configuration is different from the conventional T$_{\rm cc}$ discussed above. The new state of T$_{\rm cc}$ has the configuration of cc ($\bar{\rm u}\bar{\rm d}$) pair in color $\bm{6}$ ($\bm{\bar{6}}$) and spin $^{1}{\rm S}_{0}$ ($^{3}{\rm S}_{1}$). We find that the $\bar{\rm u}\bar{\rm d}$ pair in color $\bm{\bar{6}}$ and spin $^{3}{\rm S}_{1}$ in isospin $I=0$ is attractive, although its strength is smaller (factor $-1/8$ from Table~\ref{table:color-spin}) than that of the $\bar{\rm u}\bar{\rm d}$ diquark in color $\bm{3}$, spin $^{1}{\rm S}_{0}$ and isospin $I=0$ (factor $-3/4$). Denoting the quantum numbers of the cc pair, we shall refer to the conventional (new) state as T$_{\rm cc}[\bm{\bar{3}},{}^{3}{\rm S}_{1}]$ (T$_{\rm cc}[\bm{6},{}^{1}{\rm S}_{0}]$). The quantum number of T$_{\rm cc}[\bm{6},{}^{1}{\rm S}_{0}]$ is again $I(J^{P})=0(1^{+})$. 

\begin{figure}[tbp]
\centering
\includegraphics[width=10cm,angle=0,bb=0 50 1000 550]{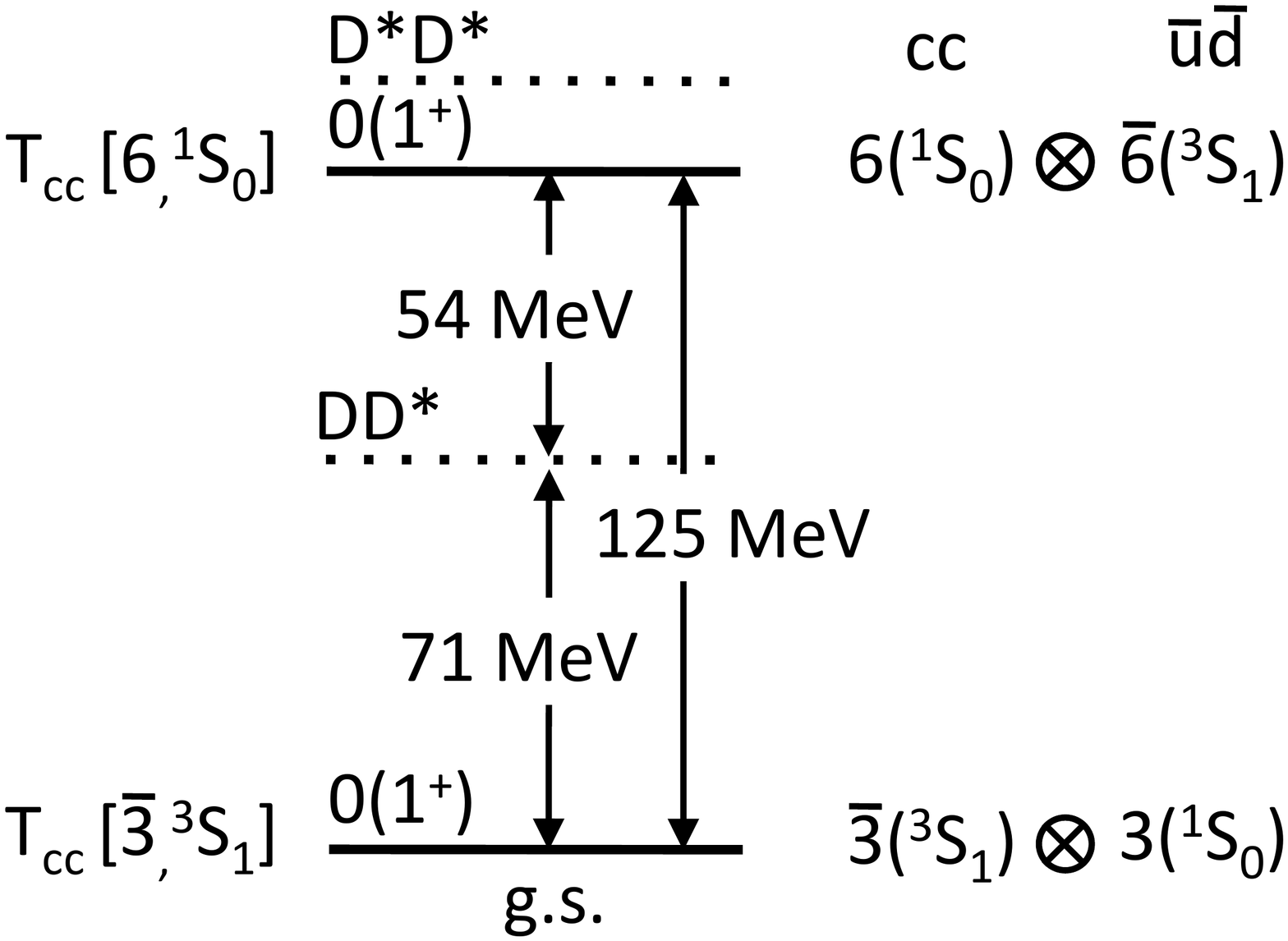}
\caption{Masses of T$_{\rm cc}[\bm{\bar{3}},{}^{3}{\rm S}_{1}]$ and T$_{\rm cc}[\bm{6},{}^{1}{\rm S}_{0}]$ (solid lines) from the diquark model. The DD$^{\ast}$ and D$^{\ast}$D$^{\ast}$ thresholds are indicated by the dotted lines. The color and spin configurations of quarks are also shown.}
\label{fig:Fig1}
\end{figure}

In principle, T$_{\rm cc}[\bm{\bar{3}},{}^{3}{\rm S}_{1}]$ and T$_{\rm cc}[\bm{6},{}^{1}{\rm S}_{0}]$ can be mixed because both of them have the same quantum numbers $I(J^{P})=0(1^{+})$. However, the spins of cc in T$_{\rm cc}[\bm{\bar{3}},{}^{3}{\rm S}_{1}]$ and T$_{\rm cc}[\bm{6},{}^{1}{\rm S}_{0}]$ are different, so the transition requires the spin-flip of a heavy quark which is suppressed by $1/m_{\rm c}$. Therefore, the mixing probability of T$_{\rm cc}[\bm{\bar{3}},{}^{3}{\rm S}_{1}]$ and T$_{\rm cc}[\bm{6},{}^{1}{\rm S}_{0}]$ is suppressed by order of $1/m_{\rm c}^2$ and we neglect the mixing effect in this work. Indeed, in a fully dynamical four-body quark-model calculation~\cite{Vijande:2009zs}, it is shown that the ground state of T$_{\rm cc}$ is dominated by the component with the cc pair in color $\bm{\bar{3}}$. We note that the suppression of the mixing is due to the spin of the heavy diquarks, not to the structure of specific interactions. The interaction without spin dependence, such as color confinement potential, cannot mix the two states, unless we consider a cc pair with odd angular momentum which will be suppressed for the ground state. We can estimate the mass splitting between T$_{\rm cc}[\bm{\bar{3}},{}^{3}{\rm S}_{1}]$ and T$_{\rm cc}[\bm{6},{}^{1}{\rm S}_{0}]$ from the color-spin interaction $H_{\rm int}$ as
\begin{eqnarray}
 M({\rm T}_{\rm cc}[\bm{6},{}^{1}{\rm S}_{0}])-M({\rm T}_{\rm cc}[\bm{\bar{3}},^{3}\!{\rm S}_{1}])\nonumber
&=&
 \left(- \frac{1}{8} \frac{C_{\rm B}}{m_{\rm u}^{2}} + \frac{3}{8} \frac{C_{\rm cc}}{m_{\rm c}^{2}}\right)
- \left( -\frac{3}{4} \frac{C_{\rm B}}{m_{\rm u}^{2}} + \frac{1}{4} \frac{C_{\rm cc}}{m_{\rm c}^{2}} \right) \nonumber \\
&\simeq& 125 \hspace{0.5em} {\rm MeV}.
\end{eqnarray}
Therefore, T$_{\rm cc}[\bm{6},{}^{1}{\rm S}_{0}]$ is an excited state of T$_{\rm cc}[\bm{\bar{3}},{}^{3}{\rm S}_{1}]$. We should remark on the stability of T$_{\rm cc}[\bm{6},^{1}{\rm S}_{0}]$. The decay of T$_{\rm cc}[\bm{6},{}^{1}{\rm S}_{0}]$ into T$_{\rm cc}[\bm{\bar{3}},{}^{3}{\rm S}_{1}]$ requires the two-pion emission at least from the isospin conservation, but this is kinematically forbidden due to the mass splitting between T$_{\rm cc}[\bm{\bar{3}},{}^{3}{\rm S}_{1}]$ and T$_{\rm cc}[\bm{6},{}^{1}{\rm S}_{0}]$ as estimated above.\footnote{The photon emission in electromagnetic interaction gives a small width.} The mass splitting indicates that the mass of T$_{\rm cc}[\bm{6},{}^{1}{\rm S}_{0}]$ can lie at 54 MeV above DD$^{\ast}$ threshold, when the mass of T$_{\rm cc}[\bm{\bar{3}},{}^{3}{\rm S}_{1}]$ lies at 71 MeV below DD$^{\ast}$ threshold (see Fig.~\ref{fig:Fig1}).
Therefore it should be kept in mind that T$_{\rm cc}[\bm{6},{}^{1}{\rm S}_{0}]$ may decay to D and D$^{\ast}$ mesons via strong interaction in s-wave fall-apart process. However, the decay to DD$^{\ast}$ requires the color recombination from the color sextet cc and antisextet $\bar{\rm u}\bar{\rm d}$ diquarks to color singlet ${\rm c}\bar{\rm u}$ and ${\rm c}\bar{\rm d}$ mesons, and the size of cc in T$_{\rm cc}[\bm{6},{}^{1}{\rm S}_{0}]$ is small because of the heavy reduced mass. These qualitative arguments suggest that T$_{\rm cc}[\bm{6},{}^{1}{\rm S}_{0}]$ can be a narrow state.

So far we have discussed the tetraquarks using $\bar{\rm u}\bar{\rm d}$ diquarks with attractive channels in Table~\ref{table:color-spin}. If we consider the other combinations for light diquarks, we may construct tetraquarks with $I(J^{P})=1(0^{+})$, $1(1^{+})$, and $1(2^{+})$. These states will have a heavier mass than the  T$_{\rm cc}[\bm{6},{}^{1}{\rm S}_{0}]$, for which the effect of the decay width may be significant. In this Letter, we concentrate on the ground states with $I=0$.

%Cross sections in NRQCD

Now let us consider the production of T$_{\rm cc}$ in inclusive processes ${\rm e}^{+}{\rm e}^{-}\to {\rm T}_{\rm cc}+{\rm X}$. We use the framework of nonrelativistic QCD (NRQCD)~\cite{Bodwin:1994jh,Petrelli:1997ge} which is an effective field theory based on the expansion in terms of the heavy quark velocity $v$. The production cross section is factorized into the short-distance perturbative amplitude and the nonperturbative matrix element of the NRQCD operators. The framework has been applied to double-charm productions with charmonia in the final states~\cite{Braaten:2002fi,Liu:2002wq,Zhang:2005cha,Bodwin:2007ga,Zhang:2008gp}. Note that the higher order corrections both in $\alpha_{s}$~\cite{Zhang:2008gp} and velocity expansion~\cite{Bodwin:2007ga} are found to be important for the double-charmonium productions. Here we provide the leading order calculation for the T$_{\rm cc}$ production as a first trial, and reserve the study of the higher order corrections for future works.

\begin{figure}[tbp]
\includegraphics[width=6.5cm,clip]{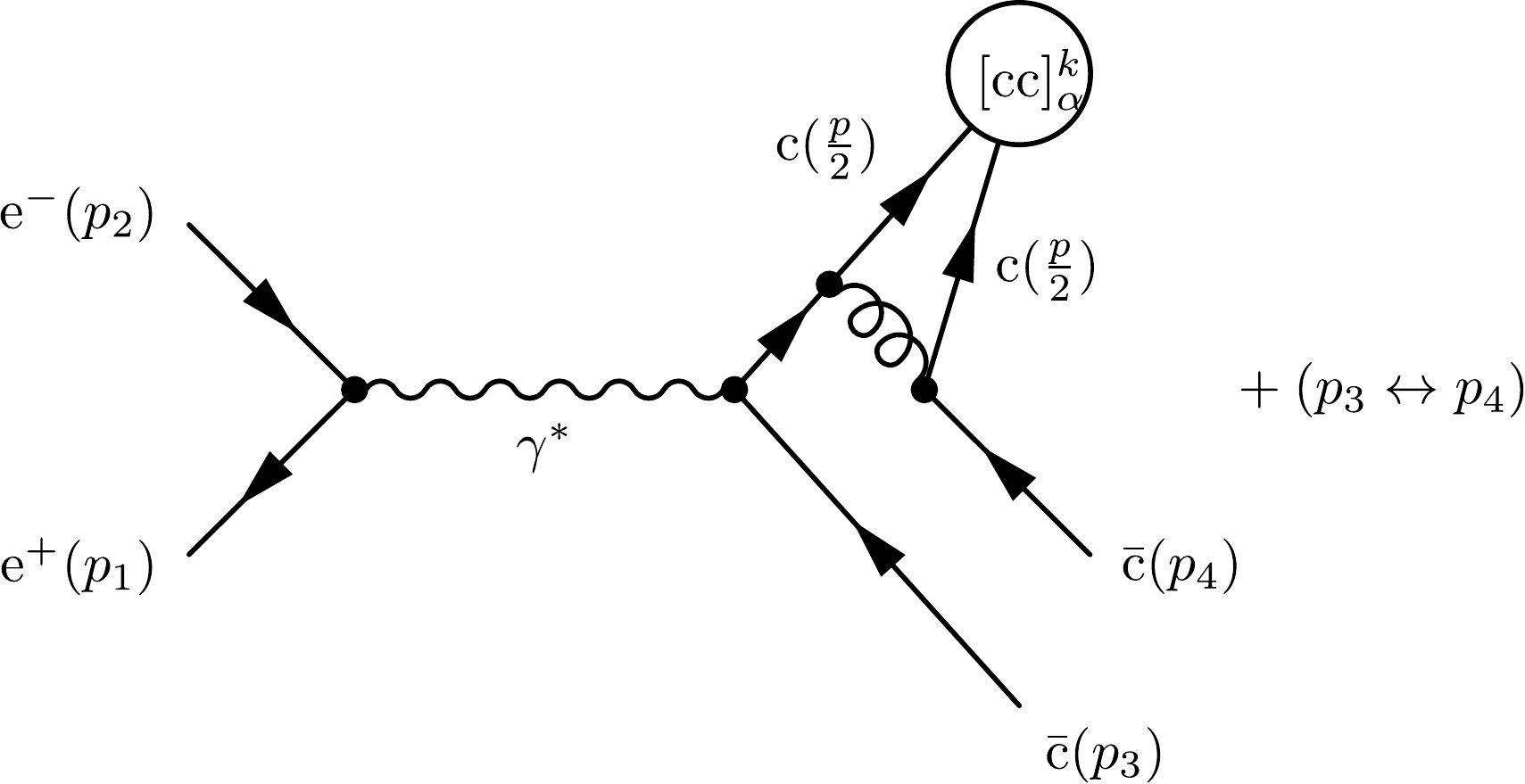} 
\includegraphics[width=6.5cm,clip]{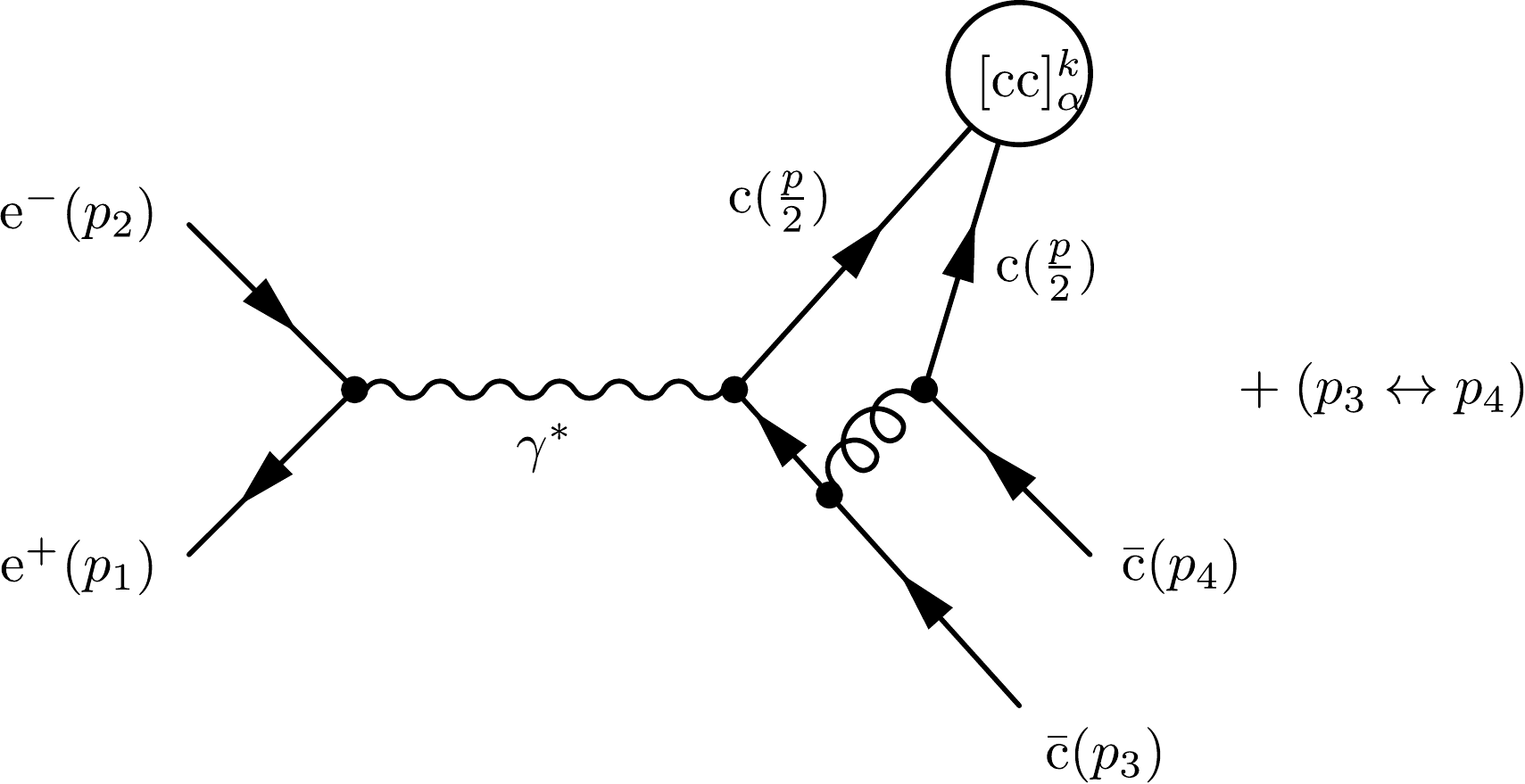}
\caption{\label{fig:Diagrams} Feynman diagrams for the leading order contributions to the perturbative ${\rm d}\hat{\sigma}({\rm e}^{+}{\rm e}^{-}\to [{\rm cc}]^{k}_{\alpha}+\bar{\rm c}+\bar{\rm c})$ process.}
\end{figure}

Here we assume that the factorization is also valid for T$_{\rm cc}$, following the strategy of Refs.~\cite{Ma:2003zk,Jiang:2012jt} where the  production of the doubly-charmed baryon $\Xi_{\rm cc}$ is studied in NRQCD. The cross section is then decomposed as 
\begin{align}
    {\rm d}\sigma_{\alpha}({\rm e}^{+}{\rm e}^{-}\to {\rm T}_{\rm cc}[\alpha]+{\rm X}) 
    =&
    \sum_{k}
    {\rm d}\hat{\sigma}({\rm e}^{+}{\rm e}^{-}\to [{\rm cc}]^{k}_{\alpha}+\bar{\rm c}+\bar{\rm c})
    \langle \mathcal{O}^{k}({\rm T}_{{\rm cc}}[\alpha])\rangle ,
    \nonumber
\end{align}
where ${\rm d}\hat{\sigma}$ represents the short-distance part of the process ${\rm e}^{+}{\rm e}^{-}\to [{\rm cc}]^{k}_{\alpha}+\bar{\rm c}+\bar{\rm c}$ with the cc pair being projected onto the definite color-spin state labeled by $\alpha= [\bm{\bar{3}},{}^{3}{\rm S}_{1}]$ or $[\bm{6},{}^{1}{\rm S}_{0}]$ for T$_{\rm cc}[\bm{\bar{3}},{}^{3}{\rm S}_{1}]$ or T$_{\rm cc}[\bm{6},{}^{1}{\rm S}_{0}]$, respectively, and $k$ specifies NRQCD operators which are sorted out by velocity expansion. Feynman diagrams for the leading order in perturbative QCD contributions to ${\rm d}\hat{\sigma}$ are shown in Fig.~\ref{fig:Diagrams}. The matrix element $\langle \mathcal{O}^{k}({\rm T}_{{\rm cc}}[\alpha])\rangle$ represents the long-distance nonperturbative process of the cc pair into T$_{\rm cc}$. In the leading order of NRQCD, the matrix element for each channel is given by a single constant as
\begin{align}
    \langle \mathcal{O}^{k}({\rm T}_{\rm cc}[\alpha])\rangle
    \Bigr|_{k={\rm LO}}
    =&
    \begin{cases}
    h_{[\bm{\bar{3}},{}^{3}{\rm S}_{1}]} & \text{for } 
    \alpha = [\bm{\bar{3}},{}^{3}{\rm S}_{1}] ,\\
    h_{[\bm{6},{}^{1}{\rm S}_{0}]} & \text{for } 
    \alpha = [\bm{6},{}^{1}{\rm S}_{0}] .
    \end{cases}
    \nonumber 
\end{align}
As we have discussed, since the different color-spin states do not mix with each other in the heavy quark limit, we calculate their productions separately.\footnote{The final state interaction of T$_{\rm cc}[\bm{6},{}^{1}{\rm S}_{0}]$ with other hadrons may affect on the production of T$_{\rm cc}[\bm{\bar{3}},{}^{3}{\rm S}_{1}]$ and vice versa. However, such a quantitative analysis is not covered in the present study and is left for future works.} In fact, the color $\bm{6}$ ($\bm{\bar{3}}$) cc component in T$_{\rm cc}[\bm{\bar{3}},{}^{1}{\rm S}_{0}]$ (T$_{\rm cc}[\bm{6},{}^{1}{\rm S}_{0}]$ ) is included in higher order in NRQCD expansion.

%--figure---------------------------------
%--figure---------------------------------

%Kinematics

To evaluate ${\rm d}\hat{\sigma}$, we assign kinematic variables as ${\rm e}^{+}(p_{1}){\rm e}^{-}(p_{2})\to [{\rm cc}]^{k}_{\alpha}(p)+\bar{\rm c}(p_{3})+\bar{\rm c}(p_{4})$. We work in the center-of-mass frame of the ${\rm e}^{+}{\rm e}^{-}$ collisions with the $z$ axis in the beam direction. Defining the $x$ axis so that the three-momentum of the produced T$_{\rm cc}$ lies in the $xz$ plane, we write the four-momentum of T$_{\rm cc}$ as
\begin{align}
    p^{\mu}
    =&
    (E_{p},p\sin\Theta,0,p\cos\Theta) ,
    \nonumber
\end{align}
where $E_{p}=\sqrt{4m_{\rm c}^{2}+p^{2}}$, $p=|\bm{p}|$ is the magnitude of the three-momentum of T$_{\rm cc}$ and $\Theta$ is the angle of $\bm{p}$ from the $z$ axis. The momenta of $\bar{c}$ are specified by $\bm{p}_{3}=-\bm{p}/2+\bm{q}$ and $\bm{p}_{4}=-\bm{p}/2-\bm{q}$, and $\bm{q}$ is expressed in the cylindrical polar coordinates as $\bm{q}=(\tilde{q}\sin\theta,q_{y},\tilde{q}\cos\theta)$.

The final expression of the differential cross section is
\begin{align}
    \frac{{\rm d}\sigma_{\alpha}}{{\rm d}p\ {\rm d}\cos\Theta}
    =&
    \frac{1}{(2\pi)^{4}}\frac{p^{2}}{16m_{\rm c}sE_{p}}
    \int_{0}^{2\pi}{\rm d}\theta
    \int_{0}^{\tilde{q}_{\text{max}}}{\rm d}\tilde{q} 
    \frac{\tilde{q}|\mathcal{M}_{\alpha}|^{2}h_{\alpha}}{q_{y}(E_{3}+E_{4})}
    \nonumber ,\\
    q_{y}
    =&
    \frac{
    \sqrt{
    A
    -B\tilde{q}^{2}
    +C\tilde{q}^{2}\cos^{2}\theta^{\prime}}}{2(\sqrt{s}-E_{p})}
    , \nonumber \\
    \tilde{q}_{\text{max}}
    =&\sqrt{\frac{A}{B-C \cos^{2}\theta^{\prime}}}
    , \nonumber 
\end{align}
where $s$ is the total energy squared, $E_{3,4}=\sqrt{m_{c}^{2}+|\bm{p}_{3,4}|^{2}}$, $\theta^{\prime}=\theta-\Theta$, $A = \sqrt{s}(\sqrt{s}-2E_{p})(\sqrt{s}-E_{p})^{2}$, $B=4(\sqrt{s}-E_{p})^{2}$, and $C=4p^{2}$. 

The amplitude for the ${\rm e}^{+}{\rm e}^{-}\to [\rm{cc}]^{k}_{\alpha}+\bar{\rm c}+\bar{\rm c}$ process $\mathcal{M}_{\alpha}$ is calculated by the diagrams in Fig.~\ref{fig:Diagrams}, with the color-spin projection
\begin{align}
    P^{(\lambda)}_{\bm{\bar{3}},m}
    =&\sum_{\bm{\bar{3}},{}^{3}{\rm S}_{1}}
    \bar{u}_{k}\left(\frac{p}{2}\right)^{t}
    \bar{u}_{j}\left(\frac{p}{2}\right)
    =\frac{1}{\sqrt{2}}
    \left(\frac{\Slash{p}}{2}+m_{\rm c}\right)
    \Slash{\epsilon}^{(\lambda)t}C\Phi_{mkj}^{\rm A}
    , \nonumber \\
    P_{\bm{6},m}
    =&\sum_{\bm{6},{}^{1}{\rm S}_{0}}
    \bar{u}_{k}\left(\frac{p}{2}\right)^{t}
    \bar{u}_{j}\left(\frac{p}{2}\right)
    =\frac{1}{\sqrt{2}}
    \left(\frac{\Slash{p}}{2}+m_{\rm c}\right)
    \gamma_{5}C\Phi_{mkj}^{\rm S}
    , \nonumber 
\end{align}
where the sum is taken for each color-spin state, $\epsilon^{(\lambda)}_{\mu}$ is the polarization vector of T$_{\rm cc}$, $C$ is the charge conjugation matrix, and $\Phi_{mkj}^{\rm A,S}=\mp\Phi_{mjk}^{\rm A,S}$ are the normalized tensor in color space. The antisymmetric part is related to the Levi-Civita symbol $\Phi_{mkj}^{\rm A}=\epsilon_{mkj}/\sqrt{2}$.

%Numerical result 1 (momentum dependence)

\begin{figure}[tbp]
\centering
\includegraphics[width=6cm,clip]{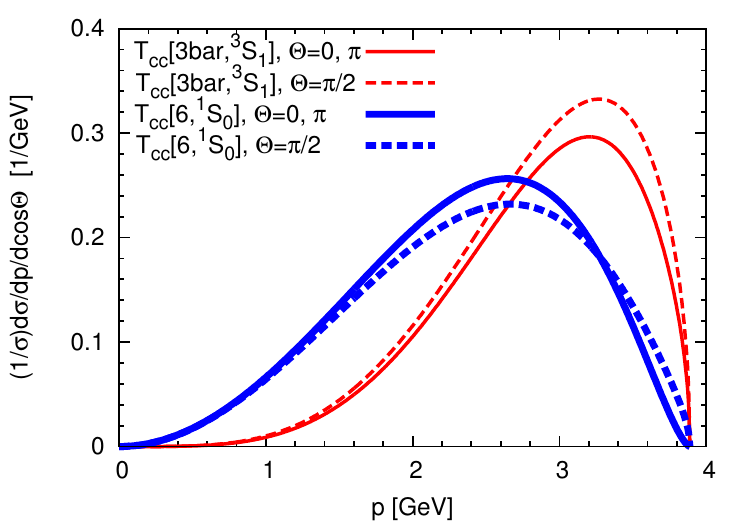}
\includegraphics[width=6cm,clip]{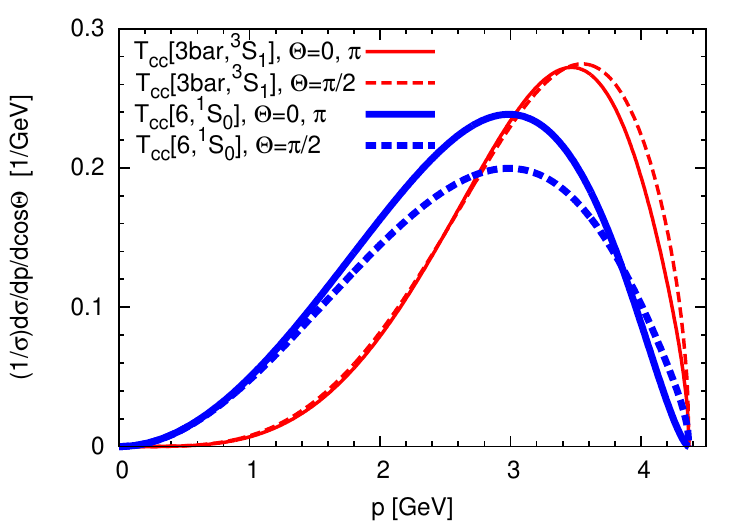}
\caption{\label{fig:result} (Color online) Differential cross sections as functions of the magnitude of the three momentum of T$_{\rm cc}$. The left (right) panel represents the results with $m_{\rm c}=1.8$ GeV and $\alpha_s = 0.212$ ($m_{\rm c}=1.5$ GeV and $\alpha_s = 0.26$). Thin (thick) lines represent the production of T$_{\rm cc}[\bm{\bar{3}},{}^{3}{\rm S}_{1}]$ (T$_{\rm cc}[\bm{6},{}^{1}{\rm S}_{0}]$).}
\end{figure}

We fix the total energy at $\sqrt{s}=10.6$ GeV as in Belle experiment~\cite{Abe:2002rb}. Following the discussion of the production of doubly charmed baryon $\Xi_{\rm cc}$ in Ref.~\cite{Jiang:2012jt}, the other constants are chosen to be $m_{\rm c}=1.8$ GeV and $\alpha_{s}=0.212$. In the leading order of NRQCD, the mass of T$_{\rm cc}$ is given by $M_{{\rm T}_{\rm cc}}=2m_{\rm c}$. Note that the mass of charm quark is different from that used in the diquark model. We calculate the differential cross sections normalized by the total cross section $(1/\sigma_{\alpha}){\rm d}\sigma_{\alpha}/({\rm d}p\ {\rm d}\cos\Theta)$ which is independent of the value of the matrix element $h_{\alpha}$. The results are plotted as functions of the T$_{\rm cc}$ momentum $p=|\bm{p}|$ in the left panel of Fig.~\ref{fig:result} with $\Theta=0$, $\pi/2$, and $\pi$. Because the amplitude is symmetric under the exchange $p_{1}\leftrightarrow p_{2}$, the result with $\Theta$ and that with $\pi-\Theta$ are identical. In Fig.~\ref{fig:result} we see that the momentum distributions of the two color states are quite different. The T$_{\rm cc}[\bm{\bar{3}},{}^{1}{\rm S}_{0}]$ channel has maximum at $p\simeq 3.5$ GeV with short tail in the low momentum region, while the distribution of the T$_{\rm cc}[\bm{6},^{1}{\rm S}_{0}]$ state peaks at around $p\simeq 2.7$ GeV with appreciable strength in the low momentum region. This qualitative difference can be used to distinguish the color structure of T$_{\rm cc}$; for instance, the ratio of the cross sections at $p=1$ GeV and $p=3.5$ GeV is quite different in two cases. 

%--figure---------------------------------
%--figure---------------------------------

% Estimate of h3 and h6 from the quark model with H.O. potential

The nonperturbative constants $h_{[\bm{\bar{3}},{}^{3}{\rm S}_{1}]}$ and $h_{[\bm{6},{}^{1}{\rm S}_{0}]}$ should in principle be estimated in QCD. In the present study, as a simple approach, we use the nonrelativistic quark model. We assume $h_{[\bm{\bar{3}},{}^{3}{\rm S}_{1}]}=|R_{\rm cc}^{{\rm T}_{\rm cc}[\bm{\bar{3}},{}^{3}{\rm S}_{1}]}(0)|^2/4\pi$ and $h_{[\bm{6},{}^{1}{\rm S}_{0}]}=|R_{\rm cc}^{{\rm T}_{\rm cc}[\bm{6},{}^{1}{\rm S}_{0}]}(0)|^2/4\pi$, respectively, where $R_{\rm cc}^{{\rm T}_{\rm cc}[\bm{\bar{3}},{}^{3}{\rm S}_{1}]}(r)$ [$R_{\rm cc}^{{\rm T}_{\rm cc}[\bm{6}_{\rm c},^{1}{\rm S}_{0}]}(r)$] is the radial wave function of ${\rm c}{\rm c}$ quark pair in tetraquark T$_{\rm cc}[\bm{\bar{3}},{}^{3}{\rm S}_{1}]$ (T$_{\rm cc}[\bm{6},{}^{1}{\rm S}_{0}]$). To obtain the wave function $R_{\rm cc}^{{\rm T}_{\rm cc}[\bm{\bar{3}},{}^{3}{\rm S}_{1}]}(r)$ and $R_{\rm cc}^{{\rm T}_{\rm cc}[\bm{6},{}^{1}{\rm S}_{0}]}(r)$, we consider the Hamiltonian with the harmonic oscillator potential $\sum_{i<j} \left( -\frac{3}{16} \right) \vec{\lambda}_{i} \!\cdot\! \vec{\lambda}_{j} \, \frac{k}{2} |\vec{r}_{i}-\vec{r}_{j}|^2$ where $\vec{r}_{i}$ is the position of the quark $i$, and $k$ is the strength parameter of the harmonic oscillator potential for quark confinement. The color factor $\vec{\lambda}_{i} \!\cdot\! \vec{\lambda}_{j}$ is important to obtain the wave function of the ${\rm c}{\rm c}$ quark pair, because the wave function should be different for each color channel, $\bm{\bar{3}}$ and $\bm{6}$. Although the harmonic oscillator potential is a simple potential in the quark model, it will be enough for our purpose when the cc wave function contains only the s wave. We note again that no mixing of $\bm{\bar{3}}$ and $\bm{6}$ is induced by the confinement interaction. The masses of quarks are set to be the same value used in Eq.~(\ref{eq:H_int}). The strength parameter $k=0.33$ GeV$^3$ is fixed to reproduce the value of the wave function at the center-of-mass for ${\rm c}\bar{\rm c}$ quark pair in charmonia, $|R_{{\rm c}\bar{\rm c}}(0)|^2$, which is estimated in the more sophisticated quark model with the Cornel-type (Coulomb + linear confinement) potential and the spin-spin interaction in Ref.~\cite{Barnes:2005pb}. Here we obtain $|R_{{\rm c}\bar{\rm c}}(0)|^2=(|R_{{\rm c}\bar{\rm c}}^{\eta_{\rm c}}(0)|^2+3|R_{{\rm c}\bar{\rm c}}^{{\rm J}/\psi}(0)|^2)/4=(1.18)^2$ GeV$^3$ with the wave functions $|R_{{\rm c}\bar{\rm c}}^{\eta_{\rm c}}(0)|^2=(1.39)^2$ GeV$^3$ and $|R_{{\rm c}\bar{\rm c}}^{{\rm J}/\psi}(0)|^2=(1.10)^2$ GeV$^3$ at the center of mass of ${\rm c}\bar{{\rm c}}$ in $\eta_{\rm c}$ and ${\rm J}/\psi$, respectively. With all the parameters being fixed, we calculate the cc wave functions in T$_{\rm cc}[\bm{\bar{3}},{}^{3}{\rm S}_{1}]$ and T$_{\rm cc}[\bm{6},{}^{1}{\rm S}_{0}]$. The harmonic oscillator potential has the property that the ${\rm c}{\rm c}$ wave function is exactly decoupled from the other light quarks ($\bar{\rm u}$ and $\bar{\rm d}$). Here we may ignore the color-spin interaction for cc pair, because in first the cc wave function is decoupled already from the other light quarks, and in second the color-spin interaction for the ${\rm c}{\rm c}$ quarks is small with suppression factor by $1/m_{\rm c}^{2}$. From the harmonic oscillator potential, as numerical result, we obtain $h_{[\bm{\bar{3}},{}^{3}{\rm S}_{1}]}=0.089$ GeV$^3$ and $h_{[\bm{6}_{\rm c},{}^{1}{\rm S}_{0}]}=0.054$ GeV$^3$. By substituting these values, we finally obtain the total cross section at Belle energy as $\sigma_{[\bm{\bar{3}},{}^{3}{\rm S}_{1}]}=13.8$ fb and $\sigma_{[\bm{6},{}^{1}{\rm S}_{0}]}=4.1$ fb. We may use the other parameter set, for example $m_{\rm c}=1.5$ GeV and $\alpha_s = 0.26$ from Ref.~\cite{Liu:2002wq}. Then we find a few times larger cross sections $\sigma_{[\bm{\bar{3}},{}^{3}{\rm S}_{1}]}= 65$ fb, $\sigma_{[\bm{6},{}^{1}{\rm S}_{0}]}=21$ fb. The momentum dependence remains qualitatively unchanged as shown in the right panel of Fig.~\ref{fig:result}.

Some remarks are ready. The cross section of the doubly charmed baryon $\Xi_{\rm cc}$ has been discussed in a similar method~\cite{Ma:2003zk,Jiang:2012jt}. When the harmonic oscillator potential is used, interestingly, the wave function of cc in $\Xi_{\rm cc}$ can be estimated as the same with that of cc in the tetraquark T$_{\rm cc}[\bm{\bar{3}},{}^{3}{\rm S}_{1}]$.
Then, we obtain the cross section of $\Xi_{\rm cc}$ which is same as that of T$_{\rm cc}[\bm{\bar{3}},{}^{3}{\rm S}_{1}]$. The obtained $\Xi_{cc}$ total cross section is comparable to the previous results of $\sim 10^{1}$-$ 10^{2}$ fb~\cite{Ma:2003zk,Jiang:2012jt}. 

However, there are two issues. First, we have used a very simplified quark model having the harmonic oscillator potential. Although we expect that it will be valid to consider only the s-wave state, we will need to analyze the wave functions in more realistic potential such as the Cornel-type potential for more quantitative discussion. Second, we have assumed that the long distance quantities $h_{[\bm{\bar{3}},{}^{3}{\rm S}_{1}]}$ and $h_{[\bm{6},{}^{1}{\rm S}_{0}]}$ as well as $h_{\Xi_{\rm cc}}$ for $\Xi_{\rm cc}$ are related to the ${\rm c}{\rm c}$ wave function. It should be noted that the difference of the numbers of the light quarks (two for T$_{\rm cc}[\bm{\bar{3}},{}^{3}{\rm S}_{1}]$ and T$_{\rm cc}[\bm{6},{}^{1}{\rm S}_{0}]$, one for $\Xi_{\rm cc}$) is not accounted for in the present estimate of the cross section. However, as the number of light quarks is increased, it should be expected that the production of the hadrons becomes more suppressed. In general, the probability of picking up two light quarks is considered to be about one order of magnitude smaller than that with one light quark~\cite{Affolder:1999iq,Gelman:2002wf}. Thus, the cross section of T$_{\rm cc}[\bm{\bar{3}},{}^{3}{\rm S}_{1}]$ and T$_{\rm cc}[\bm{6},{}^{1}{\rm S}_{0}]$ should be smaller than that of $\Xi_{\rm cc}$. For more quantitative study, we need to include the light quark degrees of freedom in the fragmentation process.

In summary, we discuss the exotic color configurations in doubly charmed tetraquark T$_{\rm cc}$. We discuss that T$_{\rm cc}$ with color $\bm{6}$ configuration does not mix with conventional T$_{\rm cc}$ with color $\bm{\bar{3}}$ in the heavy quark limit. We evaluate the production cross section of T$_{\rm cc}$ in the electron-positron collisions by NRQCD at leading order both in $\alpha_{s}$ and velocity expansions. As a result we find that T$_{\rm cc}$ with color $\bm{6}$ has different momentum and angular dependence from that of T$_{\rm cc}$ with color $\bm{\bar{3}}$. This study will be useful to pin down the color configurations in exotic hadrons from experimental measurement.

\section*{Acknowledgements} 

T.H. is grateful to the support from the Global Center of Excellence Program by MEXT, Japan through the Nanoscience and Quantum Physics Project of the Tokyo Institute of Technology. 
S.Y. is supported by Grant-in-Aid for Scientific Research 
on Priority Areas ``Elucidation of New Hadrons with a Variety of Flavors 
(E01: 21105006)".
This work was partly supported by the Grant-in-Aid for Scientific Research from 
MEXT and JSPS (Grants No. 24105702, No. 24740152 % Hyodo 
and No. 22740174), % Sudoh
and by Independent Innovation Foundation of Shandong University. % Liu

%\bibliographystyle{h-physrev4}
%%\bibliographystyle{apsrev4-1}
%\bibliography{refs,refs05,myrefs}

\end{document}